\documentclass[preprint,aps,showpacs]{revtex4}

\begin{document}

\title{Gauge Invariant Treatment of the Energy Carried by a Gravitational
Wave}

\author{Philip D. Mannheim} 
\affiliation{Department of Physics,
University of Connecticut, Storrs, CT 06269}
\email{philip.mannheim@uconn.edu}

\date{June 23, 2006}

\begin{abstract}

Even though the energy carried by a gravitational wave is not itself gauge
invariant, the interaction with a gravitational antenna of the
gravitational wave which carries that energy is. It therefore has to be
possible to make some statements which involve the energy which are in
fact gauge invariant, and it is the objective of this paper to provide 
them. In order to develop a gauge invariant treatment of the issues
involved, we construct a specific action for gravitational fluctuations
which is gauge invariant to second perturbative order. Then, via
variation of this action, we obtain an energy-momentum tensor for
perturbative gravitational fluctuations around a general curved
background whose covariant conservation condition is also fully gauge
invariant to second order. Contraction of this energy-momentum
tensor with a Killing vector of the background conveniently allows us to
convert this covariant conservation condition into an ordinary
conservation condition which is also gauge invariant through second
order. Then, via spatial integration we are able to obtain a
relation involving the time derivative of the total energy of the
fluctuation and its asymptotic spatial momentum flux which is also
completely gauge invariant through second order. It is only in making the
simplification of setting the asymptotic momentum flux to zero that one
would actually lose manifest gauge invariance, with only invariance under
those particular gauge transformations which leave the asymptotic
momentum flux zero then remaining. However, if one works in an arbitrary
gauge where the asymptotic momentum flux is non-zero, the gravitational
wave will then deliver both energy and momentum to a gravitational
antenna in a completely gauge invariant manner, no matter how badly
behaved at infinity the gauge function might be.

\end{abstract}

\pacs{95.30.Sf,04.20.-q,04.30.-w}

\maketitle

\section{Introduction}

In standard treatments of the energy carried by a gravitational
fluctuation, the use of a non-covariant energy-momentum pseudo-tensor
totally obscures the covariance and gauge issues involved, while
additionally forcing one to only admit those particular gauge
transformations which are are asymptotically flat. However, the full
gauge invariance of general relativity equally holds for asymptotically
badly-behaved gauge transformations as well, with the response of a
gravitational antenna to a gravitational wave needing to be invariant
under all gauge transformations both well- or badly-behaved if such a
response is to be physically meaningful. Consequently, it is necessary to
provide a treatment of gravitational fluctuations which takes the
badly-behaved gauge transformations into account as well. Thus despite
the fact that there is no generally covariant description of the energy
carried by a gravitational wave (and not even one which is good to second
perturbative order), one still has to be able to make gauge invariant
statements regarding the interaction of the gravitational wave with a
gravitational antenna and the energy which the gravitational wave
transmits to it. It is thus the objective of this paper to provide a gauge
invariant treatment of the issues involved, using an approach which
retains full gauge invariance to second order at every step of the way.
In particular, we develop a new technique (one based on a particularly
chosen gauge-invariant action for gravitational fluctuations) for
constructing the energy-momentum tensor associated with a second order
gravitational fluctuation. And even though the particular energy-momentum
tensor we construct will not itself prove to be gauge invariant, because
of the gauge invariance of our chosen fluctuation action, its covariant
derivative nonetheless will be. However, the total energy and momentum
carried by the gravitational wave are associated not with the fluctuation
energy-momentum tensor itself but rather with the spatial integrals of
its derivatives. Consequently, the gauge invariance of the covariant
conservation condition for the energy-momentum tensor is all that we need
in order to be able to obtain an integral relation which involves both
the total energy and the total momentum of the gravitational wave which
itself is fully gauge invariant through second order. The key point of
this paper is thus that in order to secure the gauge invariance of the
integral relations one does not actually need the gauge invariance of the
fluctuation energy-momentum tensor itself but rather only that of its
covariant conservation condition. Finally, once one has secured the gauge
invariance of the integral relations, while one can then choose to work
in a gauge in which the asymptotic momentum flux actually vanishes, the
utility of our work is that we can instead go to some other (typically
asymptotically badly-behaved) gauge in which the asymptotic momentum flux
does not then vanish. In such a badly-behaved gauge the energy will then
readjust since the relation between its time derivative and the
asymptotic momentum flux is fully gauge invariant, with the gravitational
wave then delivering both energy and momentum to a gravitational antenna
in a fully gauge invariant manner.\cite{footnote0} 

\section{Fluctuations in first order}

If we start off knowing only that there is some general Einstein tensor
$G^{\mu\nu}=R^{\mu\nu}-(1/2)g^{\mu\nu}g^{\alpha\beta}R_{\alpha\beta}$ and
some general energy-momentum tensor $T^{\mu\nu}$ both of which are
independently covariantly conserved with respect to an arbitrary
gravitational metric $g_{\mu\nu}$ (i.e. on non-stationary gravitational
paths which are not required to obey the Einstein equations), the quantity
$\Delta^{\mu\nu}= G^{\mu\nu}-\kappa_4^2T^{\mu\nu}$ will then be
covariantly conserved even for gravitational paths which do not obey
$\Delta^{\mu\nu}=0$. If we now break up all these various
tensors into zeroth and first order parts so that the metric can be
written as
$g_{\mu\nu}=
g^{(0)}_{\mu\nu}+g^{(1)}_{\mu\nu}=g^{(0)}_{\mu\nu}+h_{\mu\nu}$,
$g^{\mu\nu}=g^{(0)\mu\nu}-h^{\mu\nu}$ (we use $h_{\mu\nu}$ to denote
$g^{(1)}_{\mu\nu}$), the covariant conservation of the zeroth order
$\Delta^{(0)\mu\nu}$ with respect to
$g^{(0)}_{\mu\nu}$ (which we shall require) and the covariant
conservation of the full
$\Delta^{\mu\nu}$ with respect to the full
$g_{\mu\nu}$ will then entail that the first order $\Delta^{(1)\mu\nu}$
as defined as
\begin{eqnarray}
\Delta^{(1)}_{\mu\nu}=R^{(1)}_{\mu\nu}
-{1 \over
2}h_{\mu\nu}g^{(0)\alpha\beta}R^{(0)}_{\alpha\beta}
-{1 \over 2}g^{(0)}_{\mu\nu}g^{(0) \alpha
\beta}R^{(1)}_{\alpha
\beta}
+{1 \over 2}g^{(0)}_{\mu\nu}h^{\alpha \beta} R^{(0)}_{ \alpha
\beta}- \kappa_4^2T^{(1)}_{\mu\nu}
\label{1}
\end{eqnarray}
will then obey
\begin{eqnarray}
\partial_{\nu}\Delta^{(1)
\mu\nu}+\Delta^{(1)\mu\lambda}\Gamma^{(0)\nu}_{\phantom{(1)}\nu\lambda}
+\Delta^{(1)\nu\lambda}\Gamma^{(0)\mu}_{\phantom{(1)}\nu\lambda}
+\Delta^{(0)\mu\lambda}\Gamma^{(1)\nu}_{\phantom{(1)}\nu\lambda}
+\Delta^{(0)\nu\lambda}\Gamma^{(1)\mu}_{\phantom{(1)}\nu\lambda}=0~~,
\label{2}
\end{eqnarray}
where $\Gamma^{(1)\mu}_{\phantom{(1)}\nu\lambda}$ and $R^{(1)}_{\mu\nu}$
are given by $\Gamma^{(1)\mu}_{\phantom{(1)}\nu\lambda} =
(1/2)g^{(0)\mu\rho}
\left(\nabla_{\lambda}h_{\rho\nu}+\nabla_{\nu}h_{\rho\lambda}
-\nabla_{\rho}h_{\nu\lambda}\right)$ and $R^{(1)}_{\mu\nu}
=(1/2)\left(\nabla_{\mu}\nabla_{\nu}h
-\nabla_{\alpha}\nabla_{\mu}h^{\alpha}_{\phantom{\alpha}\nu}
-\nabla_{\alpha}\nabla_{\nu}h^{\alpha}_{\phantom{\alpha}\mu}
+\nabla_{\alpha}\nabla^{\alpha}h_{\mu\nu} \right)$, and where the
covariant
$\nabla_{\mu}$ derivatives are evaluated with respect
to the zeroth order metric
$g^{(0)}_{\mu\nu}$. For stationary zeroth order paths which obey a zeroth
order Einstein equation
$\Delta^{(0)}_{\mu\nu}=0$, it then follows that all first order paths will
obey 
\begin{equation}
\nabla_{\nu}\Delta^{(1) \mu\nu}=0
\label{3}
\end{equation}
even without the imposition of any equation of motion for the first order
$h_{\mu\nu}$. As introduced, the quantity $\Delta^{(1)
\mu\nu}$ transforms as a true tensor with
respect to the zeroth order $g^{(0)}_{\mu\nu}$, and remains unchanged if
$h_{\mu\nu}$ is replaced by
$\bar{h}_{\mu\nu}=h_{\mu\nu}+\nabla_{\mu}\epsilon_{\nu} +
\nabla_{\nu}\epsilon_{\mu}$, with the $\nabla_{\nu}\Delta^{(1)
\mu\nu}=0$ condition thus being gauge invariant
to first order in $\epsilon_{\mu}$.

The value of breaking $\Delta^{\mu\nu}$ into zeroth and first order
paths, is that if we perturb the zeroth order background
$g^{(0)}_{\mu\nu}$ with some first order perturbation
$\tau^{(1)}_{\mu\nu}$ which is also conserved with respect to the
background, the  perturbation will induce changes in both the background
Einstein tensor and the background energy-momentum tensor, with the first
order
$h_{\mu\nu}$ then being fixed as the solution to
\begin{equation}
\Delta^{(1)}_{\mu\nu}=-\kappa_4^2\tau^{(1)}_{\mu\nu}
\label{4}
\end{equation}
once the first order Einstein equations are imposed.
As such, Eq. (\ref{4}) is automatically fully gauge
invariant to first order in $\epsilon_{\mu}$. 

\section{Fluctuations in second order} 

The presence of the perturbation will also lead to a second order
effect, namely the emission of a gravitational wave, and Weinberg
\cite{Weinberg1972} has suggested that we identify its energy-momentum
tensor as ($1/\kappa_4^2$ times) that part, viz. 
$\Delta^{(2)}_{\mu\nu}(h)$, of the full
$\Delta_{\mu\nu}$ which is second order in $h_{\mu\nu}$.
Since the Einstein equations take the form
\begin{equation}
\Delta^{(0)}_{\mu\nu}+\Delta^{(1)}_{\mu\nu}+\Delta^{(2)}_{\mu\nu}
=-\kappa_4^2\tau^{(1)}_{\mu\nu}
\label{5}
\end{equation}
through second order, in solutions to Eq. (\ref{5}) which obey both
$\Delta^{(0)}_{\mu\nu}=0$ and
$\Delta^{(1)}_{\mu\nu}=-\kappa_4^2\tau^{(1)}_{\mu\nu}$, the second order
$\Delta^{(2)}_{\mu\nu}$ and $\nabla_{\nu}\Delta^{(2) \mu\nu}$ (as
evaluated with respect to the background
$g^{(0)}_{\mu\nu}$) would both have to vanish identically. Since the full
second order $\Delta^{(2)}_{\mu\nu}$ does vanish, the only way for that
the piece of it which is second order in $h_{\mu\nu}$ to not itself
vanish when $h_{\mu\nu}$ is itself a solution to the first order Eq.
(\ref{4}) is if in addition to $\Delta^{(2)}_{\mu\nu}(h)$, the full
$\Delta^{(2)}_{\mu\nu}$ contains some other, intrinsically
second order, term [to be labelled
$\Delta^{(2)}_{\mu\nu}(g^{(2)})$] which would have to be equal to
$-\Delta^{(2)}_{\mu\nu}(h)$. However, in that case it would only
be the conservation of the sum of 
$\Delta^{(2)}_{\mu\nu}(h)$ and $\Delta^{(2)}_{\mu\nu}(g^{(2)})$
which would be secured by the imposition of Eq. (\ref{5}), to thus
not immediately ensure that $\Delta^{(2)}_{\mu\nu}(h)$ itself would
in fact be able to serve as a conserved gravitational wave energy-momentum
tensor. However, as we now show, on explicitly constructing the additional
$\Delta^{(2)}_{\mu\nu}(g^{(2)})$ term in the explicit case of fluctuations
around a flat background, we will find it to be conserved all on its own,
so that it does not in fact exchange energy and momentum with
$\Delta^{(2)}_{\mu\nu}(h)$, to thereby allow
$\Delta^{(2)}_{\mu\nu}(h)$ to be independently conserved after all.
Then, guided by this decoupling of the $h_{\mu\nu}$ and
$g^{(2)}_{\mu\nu}$ sectors in the flat background case, using an action
principle we shall then  generalize the decoupling to
general curved backgrounds as well.\cite{footnote00}

When we perturb a system with a first order perturbation
$\tau^{(1)}_{\mu\nu}$ we not only induce a first order change in the
metric, we will also induce higher order changes in it as well. To second
order then we must take the perturbed metric to be of the form
\begin{equation}
g_{\mu\nu}=g^{(0)}_{\mu\nu}+h_{\mu\nu}+g^{(2)}_{\mu\nu}~~.
\label{6}
\end{equation}
Through second order the associated inverse metric and determinant are
given by
\begin{equation}
g^{\mu\nu}=g^{(0)\mu\nu}-h^{\mu\nu}
+h^{\mu}_{\phantom{\mu}\sigma}h^{\sigma\nu}
-g^{(2)\mu\nu}~~,~~g=g^{(0)}\left(1+h+
\frac{h^2}{2}-\frac{h_{\mu\nu}h^{\mu\nu}}{2}
+g^{(0)\mu\nu}g^{(2)}_{\mu\nu}\right)~~,
\label{7}
\end{equation}
with the second order term in the Einstein tensor being given by
\begin{eqnarray}
G^{(2)}_{\mu\nu}&=&R^{(2)}_{\mu\nu}-{1 \over 2}g^{(0)}_{\mu\nu}
\bigg{(}g^{(0)\alpha\beta}R^{(2)}_{\alpha\beta}
-h^{\alpha\beta}R^{(1)}_{\alpha\beta}
+h^{\alpha}_{\phantom{\alpha}\sigma}h^{\sigma\beta}R^{(0)}_{\alpha\beta}
-g^{(2)\alpha\beta}R^{(0)}_{\alpha\beta}\bigg{)}
\nonumber\\
&&
-{1 \over
2}h_{\mu\nu}\bigg{(}g^{(0)\alpha\beta}R^{(1)}_{\alpha\beta}
-h^{\alpha\beta}R^{(0)}_{\alpha\beta}\bigg{)}
-{1 \over 2}g^{(2)}_{\mu\nu}g^{(0)\alpha\beta}R^{(0)}_{\alpha\beta} ~~.
\label{8}
\end{eqnarray}
To identify the specific role played by the intrinsically second order
$g^{(2)}_{\mu\nu}$ it is sufficient to descend to the flat background
case where $g^{(0)}_{\mu\nu}=\eta_{\mu\nu}$. On recalling that the general
curved space Riemann tensor is given by
$R_{\lambda\mu\nu\kappa}=(1/2)(\partial_{\kappa}\partial_{\mu}
g_{\lambda\nu}  -\partial_{\kappa}\partial_{\lambda} g_{\mu\nu}  
-\partial_{\nu}\partial_{\mu}g_{\lambda\kappa} 
+\partial_{\nu}\partial_{\lambda}g_{\mu\kappa})
+g_{\eta\sigma}(\Gamma^{\eta}_{\nu\lambda}\Gamma^{\sigma}_{\mu\kappa}
-\Gamma^{\eta}_{\kappa\lambda}\Gamma^{\sigma}_{\mu\nu})$, in the flat
background case we see that the $g^{(2)}_{\mu\nu}$ dependent term in
Eq. (\ref{8}) is given by 
\begin{eqnarray}
G^{(2)}_{\mu\nu}(g^{(2)})&=&
{1 \over 2}\bigg{(}\partial_{\mu}\partial_{\nu} 
g^{(2)\alpha}_{\phantom{(2)\alpha}\alpha}
-\partial_{\alpha}\partial_{\mu}
g^{(2)\alpha}_{\phantom{(2)\alpha}\nu}
-\partial_{\alpha}\partial
_{\nu}g^{(2)\alpha}_{\phantom{(2)\alpha}\mu}
+\partial_{\alpha}\partial^{\alpha}g^{(2)}_{\mu\nu}
\bigg{)}
\nonumber \\
&&-{1 \over 2}\eta_{\mu\nu}\bigg{(}
\partial_{\alpha}\partial^{\alpha}g^{(2)\beta}_{\phantom{(2)\beta}\beta} 
-\partial_{\alpha}\partial_{\beta}g^{(2)\alpha\beta} \bigg{)}~~.
\label{9}
\end{eqnarray}
As such this expression is quite remarkable. Specifically it says that
the dependence of the second order $G^{(2)}_{\mu\nu}(g^{(2)})$ on
$g^{(2)}_{\mu\nu}$ is precisely the same as the dependence of the first
order $G^{(1)}_{\mu\nu}$ on
$h_{\mu\nu}$ [viz. $G^{(1)}_{\mu\nu}=(1/2)(\partial_{\mu}\partial_{\nu}h
-\partial_{\alpha}\partial_{\mu}h^{\alpha}_{\phantom{\alpha}\nu}
-\partial_{\alpha}\partial_{\nu}h^{\alpha}_{\phantom{\alpha}\mu}
+\partial_{\alpha}\partial^{\alpha}h_{\mu\nu})-(1/2)\eta_{\mu\nu}
(\partial_{\alpha}\partial^{\alpha}h
-\partial_{\alpha}\partial_{\beta}h^{\alpha\beta})$] in the same flat
background. However, since
$G^{(1)}_{\mu\nu}$ kinematically obeys a linearized Bianchi identity
without any need to impose any equation of motion, it follows that
$G^{(2)}_{\mu\nu}(g^{(2)})$ must do so too, and thus we conclude that the
condition $\partial_{\nu}G^{(2)\mu\nu}(g^{(2)})=0$ not only holds, but
that it does so without needing to impose any stationarity condition on
$g^{(2)}_{\mu\nu}$ whatsoever. If however, we now do impose Eq.
(\ref{5}), we will then find that $G^{(2)}_{\mu\nu}(h)$ (and thus
$\Delta^{(2)}_{\mu\nu}(h)$ in the flat background case) will be
conserved also. After the fact then we conclude that in the flat
background case we can set 
\begin{equation}
\partial_{\nu}\Delta^{(2)\mu\nu}(h)=0
\label{10}
\end{equation}
after all, just as we want. 

An additional feature of the form of Eq.
(\ref{9}) is that once we have fixed $h_{\mu\nu}$ from the first order Eq.
(\ref{4}), the vanishing of the full $\Delta^{(2)}_{\mu\nu}$ in solutions
to Eq. (\ref{5}) would then enable us to determine $g^{(2)}_{\mu\nu}$ as a
closed form function which would indeed be quadratic in $h_{\mu\nu}$,
just as it should be. And not only that, from the explicit form of Eq.
(\ref{9}), we see that the equation for $g^{(2)}_{\mu\nu}$ would be in
the form of none other than an Einstein equation whose source
term is
$\Delta^{(2)}_{\mu\nu}(h)$. Finally, with the change in the second
order $g^{(2)}_{\mu\nu}$ under an infinitesimal gauge transformation
$x^{\mu} \rightarrow x^{\mu}-\epsilon^{\mu}$ being given by 
$g^{(2)}_{\mu\nu}\rightarrow g^{(2)}_{\mu\nu} +
h_{\mu\lambda}(x)\partial_{\nu}
\epsilon^{\lambda} +h_{\nu\lambda}(x)\partial_{\mu} \epsilon^{\lambda}
+\epsilon^{\lambda}\partial_{\lambda}h_{\mu\nu}(x)
+\partial_{\mu}(\epsilon^{\lambda}\partial_{\lambda}\epsilon_{\nu})
+\partial_{\nu}(\epsilon^{\lambda}\partial_{\lambda}\epsilon_{\mu}) 
+\partial_{\mu}\epsilon_{\lambda}\partial_{\nu}\epsilon^{\lambda}$, we
infer that the vanishing of $\partial_{\nu}\Delta^{(2)\mu\nu}(g^{(2)})$
is itself gauge invariant, with the vanishing of 
$\partial_{\nu}\Delta^{(2)\mu\nu}(h)$ then being gauge invariant
through second order too. Thus even though
$\Delta^{(2)\mu\nu}(h)$ is not itself gauge invariant, its
derivative is, with the associated integral condition
\begin{equation}
{\partial \over \partial t}\int
d^3x \Delta^{(2)00}(h)= -\int dSn_i\Delta^{(2)0i}(h)~~,
\label{11}
\end{equation}
then being gauge invariant too \cite{footnote1}. Without any loss of
gauge invariance one can thus arrive at an integral relation which
relates the time derivative of the energy to an asymptotic momentum flux,
no matter how badly behaved a gauge one might choose to work in. The only
place where gauge invariance could be lost would be in dropping the
asymptotic momentum flux term, as its vanishing does not occur in
asymptotically badly-behaved gauges. Nonetheless, in gauges where the
asymptotic momentum flux is non-vanishing, the gravitational wave would
deliver not just energy but momentum also to a gravitational antenna,
doing so in a completely gauge invariant manner \cite{footnote2}.

\section{The Reason for the Decoupling}

To understand and to then be able to generalize the above found decoupling
of the two second order sectors, we recall a very useful property of the
Einstein-Hilbert action $I_{\rm EH}=-(1/2\kappa_4^2)\int d^4x (-g)^{1/2}
R^{\alpha}_{\phantom{\alpha}\alpha}$, namely that under integration by
parts it can brought to the form
\cite{Bak1994}
\begin{equation} 
I_{\rm EH}={1 \over
2\kappa_4^2}\int (-g)^{1/2}g^{\mu\nu}
\left(\Gamma^{\alpha}_{\mu\beta}\Gamma^{\beta}_{\nu\alpha}-
\Gamma^{\alpha}_{\mu\nu}\Gamma^{\beta}_{\alpha\beta}\right)~~.
\label{12}
\end{equation}
For our purposes here the great utility of Eq. (\ref{12}) is that for a
flat background an expansion of $I_{\rm EH}$ through second order can only
involve terms which are no higher than first order in the Christoffel
symbols, to thus involve $h_{\mu\nu}$ but not $g^{(2)}_{\mu\nu}$ at all.
The entire dependence of $R^{(2)}_{\alpha\beta}$ on
$g^{(2)}_{\mu\nu}$ can thus be removed from the second order
$I^{(2)}_{\rm EH}$ by an integration by parts which would then put the
$g^{(2)}_{\mu\nu}$ dependence entirely in irrelevant surface terms.
Further, on explicitly evaluating  Eq. (\ref{12}) in a flat background,
$I^{(2)}_{\rm EH}$ is found to take the form
\begin{eqnarray} 
I^{(2)}_{\rm EH}&=&
{1 \over 8\kappa_4^2}\int d^4x
h^{\mu\nu}
(\partial_{\mu}\partial_{\nu}h
-\partial_{\alpha}\partial_{\mu}h^{\alpha}_{\phantom{\alpha}\nu}
-\partial_{\alpha}\partial_{\nu}h^{\alpha}_{\phantom{\alpha}\mu}
\nonumber \\
&&+\partial_{\alpha}\partial^{\alpha}h_{\mu\nu}
-\eta_{\mu\nu}\partial_{\alpha}\partial^{\alpha}h
+\eta_{\mu\nu}\partial_{\alpha}\partial_{\beta}h^{\alpha\beta})
~~.
\label{13}
\end{eqnarray}
We recognize Eq. (\ref{13}) to be of none other than the form 
\begin{equation} 
I^{(2)}_{\rm EH}=
{1 \over 4\kappa_4^2}\int d^4x
h^{\mu\nu}
G^{(1)}_{\mu\nu}~~,
\label{14}
\end{equation}
where $G^{(1)}_{\mu\nu}=R^{(1)}_{\mu\nu}-(1/2)\eta_{\mu\nu}R^{(1)}$ is the
first order change in the Einstein tensor in the flat background. As
such, the stationary variation of Eq. (\ref{14}) with respect to the
fluctuation $h_{\mu\nu}$ would thus yield the source free region version
of Eq. (\ref{4}) as evaluated in a flat background, viz. the first order
$G^{(1)}_{\mu\nu}=0$. As an action, the second order $I^{(2)}_{\rm EH}$ is
gauge invariant under $h_{\mu\nu}\rightarrow h_{\mu\nu} +
\partial_{\mu}\epsilon_{\nu}+\partial_{\nu}\epsilon_{\mu}$ since
$G^{(1)}_{\mu\nu}$ is itself gauge invariant and 
$\partial_{\mu}G^{(1)\mu\nu}$ is kinematically zero, and thus has to lead
to a first order wave equation $G^{(1)}_{\mu\nu}=0$ which is gauge
invariant too. In other words, since the first order wave equation is
gauge invariant, there has to exist some second order action from
which it can be derived, an action which would itself need to be
gauge invariant to second order in $\epsilon_{\mu}$. Moreover, since the
first order wave equation cannot depend on the second
order $g^{(2)}_{\mu\nu}$, the requisite gauge invariant second order
action could not depend on
$g^{(2)}_{\mu\nu}$ either. Via Eq. (\ref{14}) then discussion of
the sector which is second order in $h_{\mu\nu}$ can thus be conducted
without reference to $g^{(2)}_{\mu\nu}$ at all.

Since the $I^{(2)}_{\rm EH}$ action provides us with an equation of
motion when we vary with respect to the fluctuation $h_{\mu\nu}$, we can
view $I^{(2)}_{\rm EH}$ as describing a field theory in which a spin
two field $h_{\mu\nu}$ propagates in some background $\eta_{\mu\nu}$. For
such a spin two field theory we can construct an energy-momentum tensor.
Specifically, in Eq. (\ref{13}) we replace the metric $\eta_{\mu\nu}$ by
a general $g_{\mu\nu}$, replace ordinary derivatives by covariant ones,
and then do a functional variation of the $I^{(2)}_{\rm EH}$ action with
respect to $g_{\mu\nu}$ to construct the rank two tensor
$t^{(2)\mu\nu}=(2/(-g)^{1/2})\delta I^{(2)}_{\rm EH}/\delta g_{\mu\nu}$,
with the flat space limit of this $t^{(2)}_{\mu\nu}$ then being the
requisite flat spacetime energy-momentum tensor associated with the
propagation of $h_{\mu\nu}$ in the flat background. Moreover, with the
action $I^{(2)}_{\rm EH}$ being a general coordinate scalar, in solutions
to the first order $h_{\mu\nu}$ wave equation the second
order $t^{(2)}_{\mu\nu}$ constructed this way will automatically be
covariantly conserved.

To actually perform the requisite variation of the action of Eqs.
(\ref{13}) and (\ref{14}), we must vary with respect to $g_{\mu\nu}$
without yet imposing the first order flat background $G^{(1)}_{\mu\nu}=0$
equation of motion for
$h_{\mu\nu}$, with a fair amount of algebra then being found to yield an
associated
$t^{(2)}_{\mu\nu}$ of the form
\begin{eqnarray}
4\kappa_4^2t^{(2)}_{\mu\nu}&=&
 h^{\alpha\beta}\partial_{\alpha}\partial_{\mu}h_{\nu\beta}
+h^{\alpha\beta}\partial_{\alpha}\partial_{\nu}h_{\mu\beta}
+h_{\mu\alpha}\partial_{\nu}\partial_{\beta}h^{\alpha\beta}
+h_{\nu\alpha}\partial_{\mu}\partial_{\beta}h^{\alpha\beta}
-2\eta_{\mu\nu}h^{\alpha\beta}\partial_{\alpha}\partial_{\sigma}
h^{\sigma}_{\phantom{\sigma}\beta}
\nonumber \\
&&
-h_{\mu\alpha}\partial_{\beta}\partial^{\beta}h^{\alpha}_{{\phantom
\alpha}\nu}
-h_{\nu\alpha}\partial_{\beta}\partial^{\beta}h^{\alpha}_{{\phantom
\alpha}\mu}
-h_{\mu\alpha}\partial_{\nu}\partial^{\alpha}h
-h_{\nu\alpha}\partial_{\mu}\partial^{\alpha}h
+h_{\mu\nu}\partial_{\alpha}\partial^{\alpha}h 
\nonumber \\
&&
+\eta_{\mu\nu}h^{\alpha\beta}\partial_{\alpha}\partial_{\beta}h
+\partial_{\mu}h_{\nu\alpha}
\partial_{\beta}h^{\alpha\beta}
+\partial_{\nu}h_{\mu\alpha}
\partial_{\beta}h^{\alpha\beta}
-\partial_{\mu}h^{\alpha\beta}
\partial_{\alpha}h_{\nu\beta}
-\partial_{\nu}h^{\alpha\beta}
\partial_{\alpha}h_{\mu\beta}
\nonumber \\
&&
-2\partial_{\alpha}h^{\alpha}_{\phantom{\alpha}\mu}
\partial_{\beta}h^{\beta}_{\phantom{\beta}\nu}
+2\partial_{\alpha}h_{\mu\nu}\partial_{\beta}h^{\alpha\beta}
-\eta_{\mu\nu}\partial_{\alpha}h^{\alpha}_{\phantom{\alpha}\beta}
\partial_{\sigma}h^{\beta\sigma}
+\partial_{\mu}h^{\alpha\beta}
\partial_{\nu}h_{\alpha\beta}
-\frac{1}{2}\eta_{\mu\nu}\partial_{\sigma}h_{\alpha\beta}
\partial^{\sigma}h^{\alpha\beta}
\nonumber \\
&&
+\partial_{\alpha}h^{\alpha}_{\phantom{\alpha}\mu}\partial_{\nu}h
+\partial_{\alpha}h^{\alpha}_{\phantom{\alpha}\nu}\partial_{\mu}h
-\partial_{\alpha}h_{\mu\nu}\partial^{\alpha}h
-\partial_{\mu}h\partial_{\nu}h
+\frac{1}{2}\eta_{\mu\nu}\partial_{\alpha}h
\partial^{\alpha}h
~~.
\label{15}
\end{eqnarray}
With the covariant derivative of this $t^{(2)\mu\nu}$ evaluating to
\begin{eqnarray}
4\kappa_4^2\partial_{\mu}t^{(2)\mu\nu}&=&(\partial^{\nu}h^{\mu\alpha}
-2\partial^{\mu}h^{\nu\alpha})\bigg{[}
\partial_{\beta}\partial^{\beta}h_{\mu\alpha}
-\partial_{\mu}\partial_{\beta}h^{\beta}_{\phantom{\beta}\alpha}
-\partial_{\alpha}\partial_{\beta}h^{\beta}_{\phantom{\beta}\mu}
+\partial_{\mu}\partial_{\alpha}h
\nonumber \\
&&~
-\eta_{\mu\alpha}(\partial_{\beta}\partial^{\beta}h
-\partial_{\beta}\partial_{\sigma}h^{\beta\sigma})\bigg{]}
=2(\partial^{\nu}h^{\mu\alpha}
-2\partial^{\mu}h^{\nu\alpha})G^{(1)}_{\mu\alpha}~~,
\label{16}
\end{eqnarray}
we readily confirm that for any on-shell $h_{\mu\nu}$ which then does obey
the first order equation of motion $G^{(1)}_{\mu\nu}=0$, the second
order energy-momentum tensor does indeed obey
$\partial_{\mu}t^{(2)\mu\nu}=0$, just as it should. Under a gauge
transformation of the form $h_{\mu\nu}
\rightarrow h_{\mu\nu}+\partial_{\mu}\epsilon_{\nu}
+\partial_{\nu}\epsilon_{\mu}$, we note that while $t^{(2)\mu\nu}$
itself will then acquire terms which are both linear and quadratic in
$\epsilon_{\mu}$, because of the gauge invariance of $G^{(1)}_{\mu\nu}$,
the covariant derivative of $t^{(2)\mu\nu}$ will only
acquire a term which is linear in $\epsilon_{\mu}$, viz. the term
$-4[\partial^{\mu}\partial^{\alpha}\epsilon^{\nu}]G^{(1)}_{\mu\alpha}$.
With this specific term vanishing when $G^{(1)}_{\mu\nu}$ vanishes, and
with
$G^{(1)}_{\mu\nu}$ vanishing for every
$h_{\mu\nu}+\partial_{\mu}\epsilon_{\nu} +\partial_{\nu}\epsilon_{\mu}$
if it already vanishes for any given $h_{\mu\nu}$, the on-shell
vanishing of 
$\partial_{\mu}t^{(2)\mu\nu}$ is thus seen to be fully gauge invariant to
second order in $\epsilon_{\mu}$.

Some simplification of Eq. (\ref{15}) can be obtained by working
in the convenient harmonic gauge where
$\partial_{\nu}h^{\mu\nu}-(1/2)\partial^{\mu}h=0$, with  $t^{(2)\mu\nu}$
then reducing to \cite{footnote3} 
\begin{eqnarray}
4\kappa_4^2t^{(2)}_{\mu\nu}&=&
-\partial_{\mu}h^{\alpha\beta}
\partial_{\alpha}h_{\nu\beta}
-\partial_{\nu}h^{\alpha\beta}
\partial_{\alpha}h_{\mu\beta}
+\partial_{\mu}h^{\alpha\beta}
\partial_{\nu}h_{\alpha\beta}
+\frac{1}{4}\eta_{\mu\nu}\partial_{\alpha}h
\partial^{\alpha}h
\nonumber \\
&&+h^{\alpha\beta}
\partial_{\alpha}\partial_{\mu}h_{\nu\beta}
+h^{\alpha\beta}
\partial_{\alpha}\partial_{\nu}h_{\mu\beta}
-\frac{1}{2}h_{\mu\alpha}\partial_{\nu}\partial^{\alpha}h
-\frac{1}{2}h_{\nu\alpha}\partial_{\mu}\partial^{\alpha}h
\nonumber \\
&&-\frac{1}{2}\eta_{\mu\nu}\partial_{\sigma}h_{\alpha\beta}
\partial^{\sigma}h^{\alpha\beta}
+\frac{1}{2}\partial_{\mu}h_{\nu\alpha}\partial^{\alpha}h
+\frac{1}{2}\partial_{\nu}h_{\mu\alpha}\partial^{\alpha}h
-\frac{1}{2}\partial_{\mu}h
\partial_{\nu}h
\nonumber \\
&&-h_{\mu\alpha}\partial_{\beta}\partial^{\beta}h^{\alpha}_{{\phantom
\alpha}\nu} 
-h_{\nu\alpha}\partial_{\beta}\partial^{\beta}h^{\alpha}_{{\phantom
\alpha}\mu}
+h_{\mu\nu}\partial_{\alpha}\partial^{\alpha}h~~.
\label{17}
\end{eqnarray}
With a typical box-normalized solution to the harmonic gauge wave equation
$\partial_{\alpha}\partial^{\alpha}h_{\mu\nu}=(1/2)\eta_{\mu\nu}
\partial_{\alpha}\partial^{\alpha}h$ being of the form
$h_{\mu\nu}=2\kappa_4e^{ip\cdot
x}e_{\mu\nu}(p^{\lambda})/(2p^0)^{1/2}L^{3/2}+{\rm c.c.}$ where
$p_{\mu}p^{\mu}=0$ and where the polarization tensor obeys
$p_{\nu}e^{\mu\nu}=(1/2)p^{\mu}e^{\alpha}_{\phantom{\alpha}\alpha}$, in
such solutions the asymptotic momentum flux is found to vanish, with the
on-shell fluctuation energy then being found to be given by the
time-independent
\begin{equation}
E^{(2)}=\int d^3xt^{(2) 00}=p^{0}\left[e^{\alpha\beta}e_{\alpha\beta}- {1
\over 2} (e^{\alpha}_{\phantom{\alpha}\alpha})^2\right]~~,
\label{18}
\end{equation}
just as one would want of an energy \cite{footnote4}.
Interestingly, the value obtained for $E^{(2)}$ is precisely the same as
that which would be obtained via the relevant Weinberg prescription, viz.
$(1/\kappa_4^2)\int d^3x
[R^{(2)00}-(1/2)\eta^{00}R^{(2)\alpha}_{\phantom{(2)\alpha}\alpha}]$, when
evaluated under exactly the same conditions \cite{footnote5}.

\section{General Energy-Momentum Tensor}

In order to extend the above analysis to a general curved space
background, we first need to find a fluctuation energy-momentum tensor
which is covariantly conserved, and then need to manipulate the covariant
conservation condition in a way which will yield an ordinary
conservation condition while not losing gauge invariance. As regards the
first issue, the solution is immediately at hand, since variation with
respect to
$h_{\mu\nu}$ of the fully covariant extension of Eq. (\ref{14}), viz. 
\begin{equation} 
I^{(2)}_{\rm EH}=
{1 \over 4\kappa_4^2}\int d^4x (-g)^{1/2}
h^{\mu\nu}
\Delta^{(1)}_{\mu\nu}~~,
\label{19}
\end{equation}
leads directly to $\Delta^{(1)}_{\mu\nu}=0$. With the
energy-momentum tensor $t^{(2)\mu\nu}=(2/(-g)^{1/2})\delta I^{(2)}_{\rm
EH}/\delta g_{\mu\nu}$ constructed from this $I^{(2)}_{\rm EH}$
being covariantly conserved when $\Delta^{(1)}_{\mu\nu}=0$, and with
$I^{(2)}_{\rm EH}$ being invariant under $h_{\mu\nu} \rightarrow
h_{\mu\nu}+\nabla_{\mu}\epsilon_{\nu}+\nabla_{\nu}\epsilon_{\mu}$, the
$\nabla_{\nu}t^{(2)\mu\nu} =0$ condition is thus fully gauge
invariant to second order. 

To use the covariant conservation condition to extract an ordinary one
we follow Abbot and Deser \cite{Abbott1982} and contract
the general
$t^{(2)\mu\nu}$ with a Killing vector $K_{\nu}$ of the curved
background \cite{footnote6}. With Killing vectors obeying the
antisymmetric
$\nabla_{\mu}K_{\nu}=-
\nabla_{\nu}K_{\mu}$, and with $t^{(2)\mu\nu}$ being
symmetric (our very construction of it as $(2/(-g)^{1/2})\delta
I^{(2)}_{\rm EH}/\delta g_{\mu\nu}$ obliges it to be symmetric), the
covariant conservation of the 4-vector
$J^{\mu}=t^{(2)\mu\nu}K_{\nu}$ immediately follows since
$\nabla_{\mu}J^{\mu}=[\nabla_{\mu}t^{(2)\mu\nu}]K_{\nu}
+t^{(2)\mu\nu}\nabla_{\mu}K_{\nu}=0$. Consequently, with
$\nabla_{\mu}J^{\mu}=(-g)^{-1/2}\partial_{\mu}[(-g)^{1/2}J^{\mu}]$, Eq.
(\ref{11}) generalizes to
\begin{eqnarray}
{\partial \over \partial t}\int
d^3x (-g)^{1/2}t^{(2)0\nu}K_{\nu}&=& -\int
d^3x {\partial [(-g)^{1/2}t^{(2)i\nu}K_{\nu}]\over \partial x^i}
\nonumber \\
&=&-\int
dSn_i(-g)^{1/2}t^{(2)i\nu}K_{\nu}~~,
\label{20}
\end{eqnarray}
to yield the gauge invariant integral relation we seek \cite{footnote7}.

To illustrate the utility of our formalism, we apply it to the recently
introduced $AdS_5/Z_2$ based brane world of Randall and Sundrum
\cite{Randall1999}. In a brane world with maximally 4-symmetric branes the
background geometry is taken to be of the separable form
$ds^2=dw^2+e^{2A(|w|)}q_{\mu\nu}dx^{\mu}dx^{\nu}$ where
$w$ is the fifth coordinate, $A(|w|)$ is the so-called warp factor, and
the $w$-independent $q_{\mu\nu}$ is the induced metric on the brane. One
is interested in the propagation in this background of axial gauge,
transverse-traceless tensor fluctuations which obey the first order wave
equation
$\Delta G_{MN}^{(1)}=(1/2)[\nabla_A\nabla^Ah_{MN}+2b^2h_{MN}]=0$ where
$-b^2$ is the curvature of
$AdS_5$ and $M=(0,1,2,3,5)$. Calculation of the $t^{(2)MN}$
associated with the 5-dimensional action $I_{\rm
EH}^{(2)}=(1/4\kappa_5^2)\int d^5x (-g)^{1/2}h^{MN}\Delta G_{MN}^{(1)}$ is
rather lengthy, with it being found
\cite{Mannheim2005} to take the form 
\begin{eqnarray}
4 \kappa_5^2t^{(2)MN}&=&
h^{B}_{\phantom{B}A}\nabla_B\nabla^Mh^{NA}
+h^{B}_{\phantom{B}A}\nabla_B\nabla^Nh^{M A}
+\nabla^Mh^{AB}\nabla^Nh_{AB}
-\frac{1}{2}g^{MN}\nabla^Sh^{AB}\nabla_Sh_{AB}
\nonumber \\
&&-\nabla^Mh^{AB}\nabla_Bh^{N}_{\phantom{N}A}
-\nabla^Nh^{AB}\nabla_Bh^{M}_{\phantom{M}A}
+b^2g^{MN}h^{AB}h_{AB}
+10b^2h^{MA}h^{N}_{\phantom{N}A}
\label{21}
\end{eqnarray}
on shell, with $\nabla_{M}t^{(2)MN}$ indeed being found to vanish
identically in modes which obey
$\nabla_A\nabla^Ah_{MN}+2b^2h_{MN}=0$. Separable mode solutions to the
wave equation have a dependence on
$|w|$ of the generic form $f_m(|w|)$ where $m$ is a separation constant,
so that with $K_M=(-1,0,0,0,0)$ being a timelike $AdS_5$ Killing vector,
modes with a vanishing asymptotic momentum flux $t^{(2)05}$  will then
have a time-independent energy. For the case where the induced metric on
the brane is an $M_4$ or $AdS_4$ geometry,
$t^{(2)05}$ is found to behave asymptotically as
$f_m(|w|)[f^{\prime}_m(|w|)-2A^{\prime}f_m(|w|)]$, with the vanishing of
this quantity leading to a time-independent energy whose dependence on the
fifth coordinate is given by $\int_0^{\infty}d|w|e^{-2A}f^2_m(|w|)$
\cite{footnote8}. With the energy being a bilinear function of
$h_{\mu\nu}$, we recognize the finiteness of this integral as being
none other than the normalization condition which is ordinarily used in
the brane world \cite{footnote9}, just as needed to enable us to
construct a propagator with which to integrate Eq. (\ref{4})
\cite{footnote10}, 
\cite{footnote11}.

\begin{acknowledgments}
This work grew out of a study of brane-world fluctuations in which 
the author was engaged with Dr. A. H. Guth, Dr. D. I. Kaiser and Dr. A.
Nayeri, and the author would like to thank them for their many
helpful comments.
\end{acknowledgments}


\begin{thebibliography}{}


\bibitem{footnote0} While there is a quite extensive literature on
studies of gravitational fluctuations through second order [see e.g. M. 
Bruni, S. Matarrese, S. Mollerach and S. Sonego, Class. Quant. Grav.
{\bf 14}, 2585 (1997);
K. Nakamura, Prog. Theor. Phys. {\bf 110}, 723 (2003);
K. A. Malik and D. Wands, Class. Quant. Grav. {\bf 21}, L65 (2004);
S. Rasanen, Jour. Cosmol. Astropart. Phys. {\bf 0402}, 003 (2004);
E. W. Kolb, S. Matarrese, A. Notari and A. Riotto, Phys. Rev. D {\bf
71}, 023524 (2005);
K. Nakamura, Prog. Theor. Phys. {\bf 113}, 481 (2005);
P. Martineau and R. Brandenberger, Phys. Rev. D {\bf 72}, 023507 (2005);
K. A. Malik, Jour. Cosmol. Astropart. Phys. {\bf 0511}, 005 (2005);
S. Rasanen, Class. Quant. Grav. {\bf 23}, 1823 (2006);
E. W. Kolb, S. Matarrese and A. Riotto, On cosmic acceleration without
dark energy, astro-ph/0506534;
P. Martineau and R. Brandenberger, Back-Reaction: A Cosmological Panacea, 
astro-ph/0510523;
K. Nakamura, Second-order Gauge Invariant Cosmological Perturbation
Theory: Einstein equations in terms of gauge invariant variables,
gr-qc/0605108;
T. Buchert, J. Larena and J.-M. Alimi, Correspondence between kinematical
backreaction and scalar field cosmologies -  the `morphon field',
gr-qc/0606020], these studies have by and large concentrated on either
general formal issues or on applications to the growth of
fluctuations in cosmology. The results presented in this paper
(especially the construction of second order energy-momentum tensors
such as that of  Eq. (\ref{15}) via the introduction and then variation of
second order gauge invariant actions such as that of Eq. (\ref{14}), and 
their use to obtain gauge invariant global integral relations such as that
exhibited in Eq. (\ref{11})) are all new to the literature.
 


\bibitem{Weinberg1972} S. Weinberg, {\it Gravitation and Cosmology:
Principles  and Applications of the General Theory of Relativity} 
(Wiley, New York, 1972).

\bibitem{footnote00} For an $h_{\mu\nu}$ which obeys the first order
$\Delta^{(1)}_{\mu\nu}=-\kappa_4^2\tau^{(1)}_{\mu\nu}$ in a background
which obeys $\Delta^{(0)}_{\mu\nu}=0$, direct evaluation of the second
order contribution of $h_{\mu\nu}$ to the full $\Delta^{(2)}_{\mu\nu}$,
viz. $\Delta^{(2)}_{\mu\nu}(h)$, is explicitly found to reveal that
$\Delta^{(2)}_{\mu\nu}(h)$ is both non-zero and covariantly conserved.
It was in trying to reconcile this with the fact that, as such, Eq.
(\ref{5}) itself only leads to the covariant conservation of the full
$\Delta^{(2)}_{\mu\nu}$ (and actually even to its vanishing) in solutions
to Eq. (\ref{4}) that engendered the study presented in this paper.


\bibitem{footnote1} With $\Delta^{(2)\mu\nu}(h)$ already being
second order, any change in the coordinate measure or in coordinate
derivatives due to the coordinate transformation $x^{\mu}\rightarrow
x^{\mu}-\epsilon^{\mu}$ would only affect Eq. (\ref{11}) in third order.

\bibitem{footnote2} If a gravitational antenna has a covariantly conserved
energy-momentum tensor $\hat{\tau}_{\mu\nu}$, its coupling to the
fluctuation $h_{\mu\nu}$ can be described by an action $I=\int d^4x
(-g)^{1/2}h^{\mu\nu}\hat{\tau}_{\mu\nu}$, with this action itself being
gauge invariant, since no matter how badly behaved the function
$\epsilon_{\mu}$ might be at infinity, once $\hat{\tau}_{\mu\nu}$ is
conserved, the integral $\int d^4x(-g)^{1/2}
(\nabla^{\mu}\epsilon^{\nu}+\nabla^{\nu}\epsilon^{\mu})
\hat{\tau}_{\mu\nu}$ can be reduced to a surface term which
will then not contribute to a stationary variation of the
action with respect to the fluctuation $h_{\mu\nu}$ in which the surface
term is held fixed. For both well- and badly-behaved $\epsilon^{\mu}$ the
coupling of a gravitational wave to an antenna is thus fully
gauge invariant.

\bibitem{Bak1994} D. Bak, D. Cangemi and 
R. Jackiw, Phys. Rev. D {\bf 49}, 5173 (1994).

\bibitem{footnote3} As a check on our calculation, we also obtained this
same $t^{(2)}_{\mu\nu}$ by first putting the
action of Eq. (\ref{14}) into the harmonic gauge
before doing the variation with respect to $g_{\mu\nu}$.

\bibitem{footnote4} Transformations of the form $h_{\mu\nu}
\rightarrow h_{\mu\nu}+p_{\mu}\epsilon_{\nu}+p_{\nu}\epsilon_{\mu}$ which
leave the harmonic gauge condition invariant also leave the harmonic 
gauge $E^{(2)}$ invariant, just as they should.


\bibitem{footnote5} The Weinberg prescription and our prescription for
$t^{(2)}_{\mu\nu}$ only differ by a two-index object which is not only 
itself covariantly conserved, but which in addition makes no
on-shell ($p^0=|\bar{p}|$) contribution to $E^{(2)}$ even though the
integration in Eq. (\ref{18}) is only three- rather than
four-dimensional.


\bibitem{Abbott1982} L. F. Abbott and S. Deser, Nucl. Phys. B {\bf
195}, 76 (1982).





\bibitem{footnote6} While having Killing
vectors is not mandatory for a general spacetime, all spaces of interest
in astrophysics and cosmology do have some.

\bibitem{footnote7} As with the coordinates, changes to the
Killing vector under $x^{\mu}\rightarrow x^{\mu}-\epsilon^{\mu}$ only
affect Eq. (\ref{20}) in third order.

\bibitem{Randall1999}
L. Randall and R. Sundrum, Phys. Rev. Lett. {\bf 83}, 4690 (1999).

\bibitem{Mannheim2005} P. D. Mannheim, {\it Brane-Localized Gravity}
(World Scientific, New Jersey, 2005).

\bibitem{footnote8} For the $dS_4$ brane case the analogous energy
integral is cut off at the Cauchy horizon where the relevant $e^A$
vanishes.

\bibitem{footnote9} An analogous discussion of brane-world
normalization issues can be found in O. DeWolfe, D. Z. Freedman, S. S.
Gubser and A. Karch, Phys. Rev. D {\bf 62}, 046008 (2000), though with an
energy-momentum tensor which differs from the one used here.

\bibitem{footnote10} Despite being standard,
brane-world propagators constructed via sums over normalizable modes
actually turn out to not be causal, though closely
related ones can be constructed which then are \cite{Mannheim2005}.

\bibitem{footnote11} In passing we note that while we have confined our
discussion to gravity theories based on the Einstein equations, the
approach of this paper can readily be applied to other metric
gravitational theories. For instance, for the conformal gravity theory [a
theory which has recently been advanced as an alternative to dark matter
and dark energy (P. D. Mannheim, Progress in Particle
and Nuclear Physics {\bf 56}, 340 (2006))], in the tensor
$\Delta^{\mu\nu}$ one everywhere replaces $G^{\mu\nu}$ by the equally
kinematically covariantly conserved $W^{\mu\nu}=(-g)^{-1/2}\delta
I_W/\delta g_{\mu\nu}$ where 
$I_W=\alpha_g\int d^4x
(-g)^{1/2}C_{\mu\nu\sigma\tau}C^{\mu\nu\sigma\tau}$ is the conformal
gravity action.



\end{thebibliography}
\end{document}